\documentstyle[aps,preprint,pra]{revtex}
\begin{document}
%JM begin preamble
\newcommand{\etal}{{\em et al.}\/}
\newcommand{\IP}{inner polarization}
\newcommand{\IPF}{\IP\ function}
\newcommand{\IPFs}{\IP\ functions}
%\renewcommand{\thetable}{\Roman{table}}
%\newcommand{\auth}[2]{#1 #2, }
%\newcommand{\jcite}[4]{{\it #1} {\bf #2}, #3 (#4)}
%\newcommand{\et}{ and }
%JM if ACS journal, uncomment following lines
%\newcommand{\jcite}[4]{{\it #1} {\bf #4}, {\it #2}, #3}
%\newcommand{\auth}[2]{#2, #1;}
%\newcommand{\edit}[2]{#2, #1, Ed.}
%\newcommand{\twoedit}[4]{#2, #1; #4, #3, Eds.}
%\newcommand{\inpress}[1]{{\it #1}, in press}
%\newcommand{\subm}[1]{{\it #1}, submitted}
%\newcommand{\et}{}
%\newcommand{\twoauth}[4]{#2, #1; #4, #3;}
%\newcommand{\andauth}[2]{#2, #1;}
%\newcommand{\book}[4]{{\it #1}; #2: #3, #4}
%\newcommand{\editedbook}[5]{{\it #1}; #2; #3: #4, #5}
%\newcommand{\inbook}[5]{In {\it #1}; #2; #3: #4, #5}
%\newcommand{\tbp}{to be published}
%JM if AIP journal uncomment following lines instead
\newcommand{\auth}[2]{#1 #2, }
\newcommand{\oneauth}[2]{#1 #2, }
\newcommand{\twoauth}[4]{#1 #2 and #3 #4, }
\newcommand{\andauth}[2]{and #1 #2, }
%\newcommand{\jcite}[4]{#1 {\bf #2}, #3 (#4)}
%JM temporary fix for Mol. Phys.
\newcommand{\jcite}[4]{{\it #1} {\bf #2}, #3 (#4)}
\newcommand{\et}{ and }
\newcommand{\BOOK}[4]{{\it #1} (#2, #3, #4)}
%JM if Elsevier journal uncomment following lines instead
%\newcommand{\auth}[2]{#1 #2, }
%\newcommand{\oneauth}[2]{#1 #2, }
%\newcommand{\twoauth}[4]{#1 #2 and #3 #4, }
%\newcommand{\andauth}[2]{and #1 #2, }
%\newcommand{\jcite}[4]{#1 #2 (#4) #3}
%\newcommand{\et}{ and }
%\newcommand{\erratum}[3]{\jcite{erratum}{#1}{#2}{#3}}
%\newcommand{\BOOK}[4]{{\it #1} (#2, #3, #4)}
%\newcommand{\jcite}[4]{#1 #2 (#4) #3}
\newcommand{\JCP}[3]{\jcite{J. Chem. Phys.}{#1}{#2}{#3}}
\newcommand{\jms}[3]{\jcite{J. Mol. Spectrosc.}{#1}{#2}{#3}}
\newcommand{\jmsp}[3]{\jcite{J. Mol. Spectrosc.}{#1}{#2}{#3}}
\newcommand{\theochem}[3]{\jcite{J. Mol. Struct. ({\sc theochem})}{#1}{#2}{#3}}
\newcommand{\jmstr}[3]{\jcite{J. Mol. Struct.}{#1}{#2}{#3}}
\newcommand{\cpl}[3]{\jcite{Chem. Phys. Lett.}{#1}{#2}{#3}}
\newcommand{\cp}[3]{\jcite{Chem. Phys.}{#1}{#2}{#3}}
\newcommand{\pr}[3]{\jcite{Phys. Rev.}{#1}{#2}{#3}}
\newcommand{\jpc}[3]{\jcite{J. Phys. Chem.}{#1}{#2}{#3}}
\newcommand{\jpca}[3]{\jcite{J. Phys. Chem. A}{#1}{#2}{#3}}
\newcommand{\jcc}[3]{\jcite{J. Comput. Chem.}{#1}{#2}{#3}}
\newcommand{\molphys}[3]{\jcite{Mol. Phys.}{#1}{#2}{#3}}
\newcommand{\physrev}[3]{\jcite{Phys. Rev.}{#1}{#2}{#3}}
\newcommand{\mph}[3]{\jcite{Mol. Phys.}{#1}{#2}{#3}}
\newcommand{\cpc}[3]{\jcite{Comput. Phys. Commun.}{#1}{#2}{#3}}
\newcommand{\jcsfii}[3]{\jcite{J. Chem. Soc. Faraday Trans. II}{#1}{#2}{#3}}
\newcommand{\jacs}[3]{\jcite{J. Am. Chem. Soc.}{#1}{#2}{#3}}
\newcommand{\ijqcs}[3]{\jcite{Int. J. Quantum Chem. Symp.}{#1}{#2}{#3}}
\newcommand{\ijqc}[3]{\jcite{Int. J. Quantum Chem.}{#1}{#2}{#3}}
\newcommand{\spa}[3]{\jcite{Spectrochim. Acta A}{#1}{#2}{#3}}
\newcommand{\tca}[3]{\jcite{Theor. Chem. Acc.}{#1}{#2}{#3}}
\newcommand{\tcaold}[3]{\jcite{Theor. Chim. Acta}{#1}{#2}{#3}}
\newcommand{\jpcrd}[3]{\jcite{J. Phys. Chem. Ref. Data}{#1}{#2}{#3}}
\newcommand{\APJ}[3]{\jcite{Astrophys. J.}{#1}{#2}{#3}}
\newcommand{\astast}[3]{\jcite{Astron. Astrophys.}{#1}{#2}{#3}}
\newcommand{\arpc}[3]{\jcite{Ann. Rev. Phys. Chem.}{#1}{#2}{#3}}
\newcommand{\HVSI}[1]{$\Delta H^{\circ}_{f,#1}$[Si(g)]}

%JM end preamble

\draft
\title{Accurate ab initio anharmonic force field and heat of formation
for silane}
\author{Jan M.L. Martin\thanks{Corresponding author. Email: {\tt comartin@wicc.weizmann.ac.il} } 
and Kim K. Baldridge\thanks{Permanent address: 
San Diego Supercomputer Center MC0505, University of California, San Diego,
9500 Gilman Drive, Building 109, La Jolla, CA 92093-0505, USA}
}
\address{Department of Organic Chemistry,
Kimmelman Building, Room 262,
Weizmann Institute of Science,
76100 Re\d{h}ovot, Israel. 
}

\author{Timothy J. Lee}
\address{MS230-3, NASA Ames Research Center,
Moffett Field, CA 94035-1000, USA}
\date{{\it Molecular Physics}: received April 19, 1999; 
accepted June 10, 1999}
\maketitle
\begin{abstract}
From large basis set coupled cluster calculations and a minor
empirical adjustment, an anharmonic force field for silane
has been derived that is consistently of spectroscopic quality 
($\pm$1 cm$^{-1}$ on vibrational fundamentals)
for all isotopomers of silane studied. Inner-shell polarization
functions have an appreciable effect on computed properties and
even on anharmonic corrections.
From large basis set coupled cluster calculations and extrapolations
to the infinite-basis set limit, we obtain 
TAE$_{0}$=303.80$\pm$0.18 kcal/mol,
which includes an anharmonic zero-point energy (19.59 kcal/mol),
inner-shell correlation ($-$0.36 kcal/mol), scalar relativistic
corrections ($-$0.70 kcal/mol), and atomic spin-orbit corrections
($-$0.43 kcal/mol). In combination with the recently revised \HVSI{0},
we obtain 
$\Delta H^{\circ}_{f,0}$[SiH$_{4}$(g)]=9.9$\pm$0.4 kcal/mol, in between
the two established experimental values.
\end{abstract}

\section{Introduction}

The spectroscopy and thermochemistry of the silane (SiH$_{4}$) 
molecule have aroused interest from a number of perspectives. Its
importance as a precursor for the chemical vapor deposition (CVD) of
silicon layers has been discussed at length by Allen and 
Schaefer\cite{All86}, who also review early theoretical work on the
molecule.

The spectroscopy of the tetrahedral group IV hydrides AH$_4$ 
(A=C, Si, Ge, Sn, Pb) has
been extensively studied. For a review of early work on AH$_4$ 
(A=Si, Ge, Sn) the
reader is referred to Ref.\cite{Bur90}.

A complete bibliography on experimental work on methane and its isotopomers would
be beyond the scope of this work (see Refs.\cite{ch4,carter} for detailed references):
we do note that an accurate ab initio force field was computed\cite{ch4} by a team
involving two of us. Based on this force field, a number of theoretical spectroscopic 
studies of the excited
vibrational states of CH$_4$ were recently studied: we note in particular
a full-dimenstional variational study by Carter et al.\cite{carter}, a low-order
perturbation theoretical/resonance polyad study by Venuti et al.\cite{Halonen}, and a high-order
canonical Van Vleck perturbation theory study by Wang and Sibert\cite{Sibert}.
We also note an accurate anharmonic force field on the isoelectronic NH$_4^+$ 
molecule by
two of us.\cite{nh4}

The infrared spectrum of silane, SiH$_4$, was first studied in 1935 by
Steward and Nielsen\cite{Ste35} and a set of fundamental frequencies for
the most abundant isotopomer was first obtained in 1942 by Nielsen
and coworkers.\cite{Nie42}

The isotopomers of SiH$_4$ have been the subject of considerable high-resolution
experimental work; for instance,
we note \cite{a,b} for $^{28}$SiH$_4$, $^{29}$SiH$_4$, $^{30}$SiH$_4$,
\cite{c,d,e} for $^{28}$SiH$_3$D, \cite{g,h} for $^{28}$SiHD$_3$, and 
\cite{i,j} for $^{28}$SiD$_4$. The molecule is of considerable astrophysical
interest, having been detected spectroscopically
in the atmospheres of Jupiter and
Saturn\cite{JupSat} and in the interstellar gas cloud surrounding the
carbon star IRC+10 216\cite{IRC}

Until most recently, only fairly low-resolution data\cite{f} were available for
SiH$_2$D$_2$; as the present paper was being prepared for publication,
a high-resolution study\cite{thiel} of the 
$\{\nu_3,\nu_4,\nu_5,\nu_7,\nu_9\}$ Coriolis resonance polyad appeared, 
in which assignments
were facilitated by mixed basis set CCSD(T) and MP2 calculations of the quartic
force field.

One of the interesting features of the infrared spectra of silane is their
pronounced local-mode character (e.g. \cite{Sun95}), leading to complex resonance polyads. The strongly `local' character also inspired a study of the
SiH$_4$ spectrum up to seven quanta using algebraic methods\cite{Ler95}.

In the present work, we shall report a high-quality 
quartic force field that is of constant quality for all the isotopomers
of silane. A theoretical spectroscopy study by Wang and Sibert\cite{sih4WS} 
is currently in progress on excited states
and vibrational resonance polyads of SiH$_4$ and isotopomers, using 
high-order (6th and 8th) canonical Van Vleck perturbation theory\cite{VanVleck}
and the force field reported in the present work.

Since this can be done at very little additional computational expense,
we shall also report a benchmark atomization energy and heat of formation
of SiH$_4$. 
The thermodynamic properties of silane are linked to a controversy
concerning the heat of vaporization of silicon, which is of 
fundamental importance to computational chemists since it is required
every time one attempts to directly compute the heat of formation of
any silicon compound, be it ab initio or semiempirically.
\HVSI{0} is given in the JANAF tables\cite{Jan85} as 
106.6$\pm$1.9 kcal/mol. Desai\cite{Desai}
reviewed the available data and recommended the JANAF value, but with a
reduced uncertainty of $\pm$1.0 kcal/mol. Recently, Grev and
Schaefer (GS)\cite{Gre92} found that their ab initio calculation of the TAE of
SiH$_4$, despite basis set incompleteness, was actually {\em larger} than the
value derived from the experimental heats of formation of Si($g$), H($g$),
and SiH$_4$($g$). They concluded that the
heat of vaporization of silicon should be revised upwards
to $\Delta H^\circ_{f,0}$[Si($g$)]=108.07(50) kcal/mol, a suggestion
supported by Ochterski et al.\cite{Och95}. Very recently, however, 
Collins and Grev (CG)\cite{Col98} considered the scalar relativistic
contribution to the binding energy of silane using relativistic
coupled cluster techniques within the Douglas-Kroll\cite{DK} (no-pair)
approximation, and found a contribution of -0.67 kcal/mol. This would
suggest a downward revision of the GS value of \HVSI{0} to 107.4 
kcal/mol, which is in excellent agreement with a recent redetermination
by Martin and Taylor\cite{sif4} of 107.15$\pm$0.39 kcal/mol. (This latter
value was derived by combining a benchmark ab initio calculation of the total 
atomization energy of tetrafluorosilane, TAE$_0$[SiF$_4$], with a 
very precise fluorine bomb calorimetric measurement\cite{ohare}
of $\Delta H^\circ_f$[SiF$_4$(g)].)

In addition,
it was pointed out\cite{Gre92,Col98} that the JANAF value of
$\Delta H^\circ_{f,0}$[SiH$_4$($g$)]=10.5$\pm$0.5 kcal/mol
is in fact the Gunn and Green\cite{Gun61} value of 9.5$\pm$0.5 kcal/mol 
increased
by a correction\cite{Ros52} of +1 kcal/mol for the phase transition
Si(amorphous)$\rightarrow$Si(cr). (Gunn and Green considered this
correction to be an artifact of the method of preparation and ignored it.)

Clearly, a calibration
calculation of TAE$_0$[SiH$_4$] might be desirable,
and is the secondary purpose of the present study. 
Accurate thermochemical parameters of SiH$_4$ (and other silicon
compounds) are of practical importance for the thermodynamic and
kinetic modeling of such processes as 
laser-induced chemical vapor deposition of silicon films from
silane\cite{Tamir},
the chemical vapor deposition of tungsten contacts for ULSI (ultralarge
scale integrated circuit) chips by SiH$_4$ reduction of WF$_6$
(e.g. \cite{SiH4-WF6}) and the generation of SiOxNy films by low-pressure
chemical vapor deposition from mixtures of SiH$_4$ with N$_2$O and/or
NH$_3$ \cite{SiOxNy} (e.g. as antireflective coatings\cite{optical} and
for ultrathin capacitors\cite{ultrathin-capacitors}). 
(We also mention in passing the use of silane compounds in dentistry\cite{dent}.)

While GS's
work was definitely state of the art in its time, the attainable
accuracy for this type of compound may well have gone up an order of
magnitude in the seven years since it was published: in a recent 
systematic study\cite{W2} of total atomization energies of a variety
of first-and second-row molecules for which they are precisely known,
procedures like the ones used in the present work achieved a mean 
absolute error of 0.23 kcal/mol, which dropped to 0.18 kcal/mol if only
systems well described by a single reference determinant (as is the case
with SiH$_4$) were considered. In order to ascertain the
utmost accuracy for hydrides, 
a zero-point energy including anharmonic corrections
was found to be desirable\cite{W2}: 
this is obtained as a by-product of the accurate
anharmonic force field which is the primary subject of the present
contribution.

\section{Computational methods}

All electronic structure calculations were carried out using 
MOLPRO 97\cite{molpro} running on DEC Alpha and SGI Origin computers
at the Weizmann Institute of Science.

The CCSD(T) [coupled cluster with all single and double 
substitutions (CCSD)\cite{Pur82} supplemented with a quasiperturbative
estimate of the contribution of connected triple 
excitations\cite{Rag89}] method, as implemented in MOLPRO\cite{Ham92},
was used throughout for the electronic
structure calculations on SiH$_{4}$. For the Si($^3P$) atom, we employed the 
definition of Ref.\cite{Wat93} for the open-shell CCSD(T) energy.

The calculations including only valence correlation employed the
standard Dunning cc-pV$n$Z (correlation consistent 
valence $n$-tuple zeta\cite{Dun89}) basis sets on hydrogen
and two different variants of the cc-pV$n$Z or aug-cc-pV$n$Z (augmented
cc-pV$n$Z\cite{Ken92,Woo93}) basis sets on Si. The first variant, 
cc-pV$n$Z+1,
was used in the force field calculations, and includes an additional
high-exponent $d$ function\cite{sio} to accommodate the greater part 
of the inner-shell 
polarization effect, 
which is known to be important for both energetic and geometric
properties of second-row molecules.\cite{sio,so2} The second variant,
aug-cc-pV$n$Z+2d1f\cite{so2}, includes two high-exponent $d$ functions and
a high-exponent $f$ function, with exponents determined by 
successively multiplying the highest exponent already present for 
that angular momentum by a factor of 2.5. Such a set should 
give\cite{so2} an exhaustive account of the energetic effects of
inner-shell polarization.

Calculations including inner-shell correlation (not to be confused 
with inner-shell polarization, which is an SCF-level effect) were
carried out using the Martin-Taylor\cite{hf} core correlation basis
set. Relativistic effects were determined with the same basis set and
as ACPF (averaged coupled pair functional\cite{Gda88}) expectation
values of the first-order Darwin and mass-velocity operators\cite{Cow76,Mar83}.

Optimizations were carried out by univariate polynomial interpolation.
Force constants in symmetry coordinates
were determined by recursive application of the central finite 
difference formula: the symmetry coordinates are defined in the
same way as in previous studies\cite{ch4,nh4} on the isovalent
CH$_4$ and NH$_4^+$ molecules. The vibrational analyses were performed 
using a modified version of the SPECTRO program\cite{spectro,Gaw90}
running on an IBM RS6000 workstation at NASA Ames 
and the DEC Alpha at the Weizmann institute.
The alignment conventions for the anharmonic constants of a spherical
top follow the work of Hecht\cite{Hec60} and general formulae for these
constants were taken from the paper by Hodgkinson et al.\cite{Hod83}.
Similar to 
previous work\cite{ch4,be4} on the spherical tops Be$_4$ and 
CH$_4$, 
the accuracy of the various spectroscopic constants was 
verified by applying opposite mass perturbations of $\pm$0.00001
a.m.u. to two of the hydrogen atoms, then repeating the analysis
in the asymmetric top formalism.

Finally, the reported zero-point energies include the $E_0$ term\cite{Tru91} 
(which is the polyatomic equivalent of the $a_0$ Dunham coefficient in diatomics).

\section{Results and discussion}

\subsection{Vibrational frequencies and anharmonic force field}

An overview of the basis set convergence of the computed bond distance,
harmonic frequencies, and vibrational anharmonic corrections is given 
in Table 1.

The effect of adding inner-shell polarization functions to the cc-pVTZ 
basis set is modest but significant (0.006 \AA) on the bond distance: 
the Si--H stretching frequencies, however, are affected by 20--25 
cm$^{-1}$. The bending frequencies are not seriously affected:
somewhat surprising are the fairly strong effects on the vibrational
anharmonicities (including, to a lesser extent, the bending 
frequencies).
The overall behavior is in contrast to previous observations\cite{so2} for 
SO$_{2}$ in which the inner-polarization effects on lower-order 
properties like geometry and harmonic frequencies
are very noticeable but those on anharmonicities next to
nonexistent, but is consistent with the very strong 
basis set sensitivity noted for the first three anharmonic corrections
of the first-row diatomic hydrides by Martin\cite{ch}. 

Likewise, a rather strong sensitivity with respect to
basis set improvement from VDZ+1 over VTZ+1 to VQZ+1 is seen for the
Si--H stretching frequencies and all the anharmonicities, even as the 
harmonic bending frequencies appear to be close to converged with
the VTZ+1 basis set. It appears that in 
general, basis set sensitivity of anharmonicities of A--H stretches is 
much more pronounced than that of A--B stretches. 

The effect of inner-shell correlation, while nontrivial for the purpose
of accurate calculations, is quite a bit more modest than that of
inner-shell polarization (as measured by comparing the cc-pVTZ and
cc-pVTZ+1 results), and in fact is not dissimilar to what one would
expect for a first-row molecule (e.g. CH$_{4}$ \cite{ch4}).

We will now consider computed fundamentals for the various isotopomers
of silane with our best force field, CCSD(T)/cc-pVQZ+1. All relevant 
data are collected in Table 2.

For $^{28}$SiH$_4$, $^{29}$SiH$_4$, and $^{30}$SiH$_4$, 
agreement between the computed and observed fundamentals can only
be described as excellent, with a mean absolute deviation of
2.5 cm$^{-1}$. Agreement for the completely deuterated isotopomer
$^{28}$SiD$_4$ is even better, with a mean absolute deviation of
1.9 cm$^{-1}$. For the $^{28}$SiH$_3$D isotopomer, agreement is
likewise excellent, with a mean absolute deviation of 2.1 cm$^{-1}$.
It would appear that the force field is certainly of good enough 
quality to permit assignments for the less well known isotopomers.

For $^{28}$SiHD$_3$, the only precisely known bands are the Si--H 
stretch, $\nu_1$=2187.2070(10) cm$^{-1}$ \cite{g}, and the $\nu_5$ degenerate
bend, 850.680823(10) cm$^{-1}$ \cite{h}. Meal and Wilson \cite{f}, in their
1956 low-resolution study, assigned absorptions at 1573, 1598, and 683 
cm$^{-1}$ to $\nu_2$, $\nu_4$, and $\nu_6$, respectively. Our calculations
confirm this assignment and are on average within about 2 cm$^{-1}$
of all the above bands. $\nu_3$ was not observed by Meal and Wilson, 
and these authors speculated that it coincide with the 683 cm$^{-1}$ ($\nu_6$)
peak. Our own calculations predict a splitting of about 5.7 cm$^{-1}$ 
between $\nu_3$ and $\nu_6$; B3LYP/VTZ+1 \cite{Bec93} infrared intensity
calculations suggest that both bands should be observable.
Inspection of the relevant spectrum (Fig. 3 in Ref.\cite{f}) revealed
that, at the resolution afforded by the equipment used,
meaningful resolution between $\nu_3$ and $\nu_6$
becomes essentially impossible, especially given contamination (noted by
Meal and Wilson) from a SiH$_2$D$_2$ impurity with $\nu_4$=682.5 cm$^{-1}$.

Until most recently, the only available information for SiH$_2$D$_2$ was
the Meal and Wilson work. Our calculations, like those of R\"otger
et al.\cite{thiel}, unambiguously suggest assignment of the 1601 and 1587 cm$^{-1}$
bands to $\nu_2$ and $\nu_6$, respectively, rather than the opposite
assignment proposed by Meal and Wilson. We note that $\nu_6$ is in
a very close Fermi resonance with $\nu_5+\nu_9$ (the unperturbed levels
being only about 10 cm$^{-1}$ apart), despite the fairly small 
interaction constant $k_{569}$=-20.88 cm$^{-1}$. Our calculations confirm
the assignments for all other bands aside from $\nu_1$ and $\nu_8$, which
are calculated to be within 1 cm$^{-1}$ of each other such that a meaningful
decision on whether or not to exchange $\nu_1$ and $\nu_8$ is impossible.
The Meal-Wilson empirical force field value of 844 cm$^{-1}$ for 
$\nu_5$ (which they were unable to observe) agrees well with our 
calculation as well as with the high-resolution value\cite{thiel}
of 842.38121(9) cm$^{-1}$.

Of the very recent measurements by R\"otger et al.\cite{thiel}, all five
bands in the Coriolis pentad ($\nu_3$, $\nu_4$, $\nu_5$, $\nu_7$, and
$\nu_9$) are in excellent agreement with the present calculation
(mean absolute deviation 1.1 cm$^{-1}$).

%v3=942.74106(4)
%v4=681.62394(3)
%v5=842.38121(9)
%v7=859.750104(4)
%v9=742.64029((3)

Among the sources of residual error in the quartic
force field, neglect of inner-shell correlation and
imperfections to CCSD(T) appear to be the potentially
largest. As seen in Table 1, inclusion of core correlation
increases harmonic frequencies by as much as 7 cm$^{-1}$ 
in this case. The effect of correlation beyond
CCSD(T) was seen to work in the {\em opposite} direction
for the first-row diatomic hydrides\cite{ch}; in the present work,
we have compared FCI/VDZ+1 and CCSD(T)/VDZ+1 harmonic frequencies
for the SiH diatomic in the $X~^2\Pi$ and $a~^4\Sigma^-$ states, and found 
a reduction in $\omega_{e}$ of 4 and 10 cm$^{-1}$, respectively.
(The FCI--CCSD(T) difference for $\omega_{e}$ was found in 
Ref.\cite{ch} to converge very rapidly with the basis set.)
Since FCI frequency calculations in
a reasonable-sized basis set for SiH$_4$ are simply not
a realistic option, we have taken another track.

We have assumed that the computed CCSD(T)/VQZ+1 force
field is fundamentally sound, and that any residual 
error would mostly affect the equilibrium bond distance
and the diagonal quadratic force constants. We have
then taken our quartic force field in symmetry coordinates,
substituted the computed CCSD(T)/MTcore bond distance 
(which agrees to four decimal places with the best experimental
value), and have iteratively refined the four diagonal
quadratic force constants such that the four experimental
fundamentals of $^{28}$SiH$_4$ are exactly reproduced
by our calculation. The final adjusted force field is
given in Table 3 and is available in machine-readable
format from the corresponding author.

As seen in Table 2, our computed fundamentals for
the other isotopomers with the adjusted force field
are in essentially perfect agreement with experiment
where accurate values are available. Discrepancies
arise for some modes of SiH$_2$D$_2$, SiHD$_3$,
and SiD$_4$ where only low-resolution data are
available. Particularly the discrepancy for
$\nu_2$ of SiD$_4$ is completely out of character:
the experimental difficulties involved in its 
determination\cite{i} suggest that perhaps the
experimental value may be in error.
(A discrepancy of 1.2 cm$^{-1}$ for
$\nu_2$ in SiH$_3$D is halved upon accounting for
a Fermi resonance $2\nu_8\approx\nu_2$.)
We hope that our computed force field
will stimulate further spectroscopic work on 
SiH$_4$ and may serve as a basis for studies employing
more sophisticated vibrational treatments, such as 
the variational techniques 
variational techniques very recently applied to methane\cite{carter}
or high-order canonical Van Vleck perturbation theory. As noted
in the Introduction, a study of the latter type is
already in progress.\cite{sih4WS}

\subsection{Geometry}

At the CCSD(T)/MTcore level, we compute a bond distance of
1.4734 \AA, which we know from experience\cite{cc,c2h2}
should be very close to the true value. Ohno, Matsuura, Endo,
and Hirota (OMEH1) \cite{Ohn85} estimate an experimental $r_e$
bond distance of 1.4741 \AA\ without supplying an error bar;
in a subsequent study (OMEH2)\cite{Ohn86}, the same authors, using two 
different methods, obtain 1.4734(10) \AA\ (``method I") and 
1.4707(6) \AA\ (``method II"), respectively, where uncertainties in 
parentheses are three standard deviations. The deviation between the
(diatomic approximation) ``method II" value and our present calculation
is more than an order of magnitude greater than usual for this
level of ab initio theory, while the ``method I" value agrees to
four decimal places with our calculation. (Normally, because of neglect
of correlation effects beyond CCSD(T) which have the tendency\cite{ch}
to lengthen bonds by 0.0002--0.0006 \AA, we expect our computed bond
distance to be slightly short, rather than too long.) 
The computed bond distance of
R\"otger et al., 1.4735 \AA\ at the CCSD(T)[all electron] level in a mixed basis
set which does not contain any core correlation functions, is likewise
in excellent agreement with the OMEH2 ``method I" value.

\subsection{Atomization energy of SiH$_4$}

Using a 3-point geometric extrapolation $A+B.C^{-n}$ from the
SCF/AV$n$Z+2d1f ($n$=T,Q,5) atomization energies, we find an SCF
limit component of the total atomization energy of 259.83 kcal/mol,
only marginally different from 
the directly computed SCF/AV5Z+2d1f value
of 259.82 kcal/mol and only 0.05 kcal/mol larger than the 
GS result.

The CCSD valence correlation component was extrapolated using 
the 2-point formula\cite{Hal98} $A+B/n^{3}$ from AV$n$Z+2d1f ($n$=Q,5) results;
thus we obtain a CCSD limit of 64.26 kcal/mol, which is 0.8 kcal/mol
larger than the largest basis set value (63.45 kcal/mol) and 
1.4 kcal/mol larger than the largest basis set value of GS
(62.86 kcal/mol). Using the alternative 3-point 
extrapolation\cite{l4} $A+B/(l+1/2)^{C}$ from AV$n$Z+2d1f ($n$=T,Q,5)
we obtain a somewhat smaller basis set limit of 63.92 kcal/mol;
however, as discussed in Ref.\cite{W2}, this procedure appears to
systematically underestimate basis set limits and was found\cite{c2h4tae}
to yield excellent agreement with experiment largely due to an error
compensation with neglect of scalar relativistic effects.

At 0.81 kcal/mol, the extrapolated basis set limit contribution of 
connected triple excitations is quite modest, and differs by
only 0.02 kcal/mol from the largest basis set value of 0.79 kcal/mol.
In fact, it is largely immaterial whether the extrapolation is done
from AV$n$Z+2d1f ($n$=T,Q) or from AV$n$Z+2d1f ($n$=Q,5), and we 
obtain essentially the same result for the (T) contribution as
GS (0.82 kcal/mol). This is an illustration of the
fact\cite{Hel97} that connected triple excitations generally converge
more rapidly with basis set than the CCSD correlation energy.

Adding up the two basis set limit values, we find a valence 
correlation component to TAE of 65.05 kcal/mol; given the essentially
purely single-reference character of the SiH$_{4}$ wave function there
is little doubt that the CCSD(T) limit is very close to the full CI
limit as well.

As noted by GS, the contribution of inner-shell 
correlation of SiH$_{4}$ is negative: we find -0.365 kcal/mol compared
to their -0.31 kcal/mol. The spin-orbit contribution is trivially
obtained from the Si($^3P$) atomic fine structure\cite{Moo63} 
as -0.43 kcal/mol, while our
computed scalar relativistic contribution, -0.70 kcal/mol, is 
essentially identical to the CG value. Finally, we obtain
TAE$_{e}$=323.39 kcal/mol.

The anharmonic zero-point vibrational energy (ZPVE) from our best force
field (including $E_0$) is 19.59 kcal/mol. This is very close to the
value of 19.69 kcal/mol obtained by GS as an average of estimated
fundamentals and CISD/TZ2P harmonic frequencies:
the computational effort involved in improving this
estimate by a mere 0.1 kcal/mol would therefore have been hard to
justify if the anharmonic force field would not have been required
for another purpose. Also, from past experience\cite{c2h4more}, we
know that such good agreement between rigorous anharmonic ZPVEs and
estimates cannot be taken for granted for hydrides.

Our best TAE$_e$ and ZPVE finally
lead to TAE$_{0}$=303.80 kcal/mol, to which we attach an 
error bar of about 0.18 kcal/mol based on previous experience\cite{W2}. 
This should be compared with
the GS largest basis set result of 303.03 kcal/mol
(or 302.36 kcal/mol after applying the CG scalar 
relativistic contributions) or the value derived from JANAF
heats of formation of Si(g), H(g), and SiH$_{4}$(g), 302.62
kcal/mol. 

If we consider alternative values for \HVSI{0} of 108.1$\pm$0.5 
kcal/mol (GS), 107.4$\pm$0.5 kcal/mol (applying CG
to the latter value), or 107.15$\pm$0.38
kcal/mol (Martin \& Taylor\cite{sif4}), 
we would obtain from our calculation $\Delta H^\circ_{f,0}$[SiH$_4$($g$)]
values of 10.8$\pm$0.5, 10.1$\pm$0.5, and 9.9$\pm$0.4 kcal/mol, respectively. Only the first of these values cannot be reconciled with Gunn and Green;
the very similar values derived from the Collins-Grev-Schaefer and
Martin-Taylor \HVSI{0} agree to within accumulated error bars with both
the JANAF and Gunn-Green values for the heat of formation of silane. While
our best value of 9.9$\pm$0.4 kcal/mol at first sight slightly favors
the Gunn-Green value (in which the Si($cr$)$\rightarrow$Si($amorph$)
transition enthalpy\cite{Ros52} was considered an artifact of the manner
of preparation), the difference is ``too close to call''. We contend that 
our calculated value is more reliable than either experiment.

\section{Conclusions}

From accurate ab initio calculations and a minor empirical
adjustment, a quartic force field for silane has been derived 
that is consistently of spectroscopic quality ($\pm$1 cm$^{-1}$
on vibrational fundamentals)
for all isotopomers of silane
studied here ($^{28}$SiH$_{4}$, $^{29}$SiH$_{4}$, 
$^{30}$SiH$_{4}$, $^{28}$SiH$_{3}$D, $^{28}$SiH$_{2}$D$_{2}$,
$^{28}$SiHD$_{3}$, and $^{28}$SiD$_{4}$). As in previous studies
on second-row molecules, we found that inner-shell polarization
functions have an appreciable effect on computed properties,
and for hydrides this apparently includes the vibrational anharmonicities.

From large basis set coupled cluster calculations and extrapolations
to the infinite-basis set limit, we obtain 
TAE$_{0}$=303.80$\pm$0.18 kcal/mol,
which includes an anharmonic zero-point energy (19.59 kcal/mol),
inner-shell correlation ($-$0.36 kcal/mol), scalar relativistic
corrections ($-$0.70 kcal/mol), and atomic spin-orbit corrections
($-$0.43 kcal/mol). In combination with the recently revised 
\HVSI{0}, 107.15$\pm$0.39 kcal/mol\cite{sif4}, we obtain
$\Delta H^{\circ}_{f,0}$[SiH$_{4}$(g)]=9.9$\pm$0.4 kcal/mol,
intermediate between the JANAF and Gunn-Green values of 10.5$\pm$0.5 
and 9.5$\pm$0.5 kcal/mol, respectively.

\acknowledgments

JM is a Yigal Allon Fellow, the incumbent of the Helen and Milton
A. Kimmelman Career Development Chair, and
an Honorary Research Associate (``Onderzoeksleider
in eremandaat'') of the
National Science Foundation of Belgium (NFWO/FNRS). 
KKB was a Fulbright Visiting Scholar at the Weizmann Institute
of Science (on leave of absence from SDSC) during the course of
this work.
This research
was partially supported by the Minerva Foundation, Munich, Germany.
We thanks Drs. X.-G. Wang and E. L. Sibert III (U. of Wisconsin,
Madison) for their encouragement.

\begin{table}
\caption{Basis set convergence of computed bond distance (\AA), 
harmonic frequencies (cm$^{-1}$), and anharmonic corrections 
(cm$^{-1}$) of $^{28}$SiH$_{4}$; effect on inner-shell correlation.}
\begin{tabular}{lccccccc}
	                 &   VDZ   &   VDZ+1 & VTZ     & VTZ+1   &  VQZ+1  & MTcore  & MTnocore \\
\hline
$r_e$            & 1.49076 & 1.48572 & 1.48504 & 1.47952 & 1.47872 & 1.47339 & 1.47736 \\
$\omega_1$       &  2242.0 &  2249.3 &  2225.7 &  2250.3 &  2262.7 &  2270.6 & 2264.1 \\
$\omega_2$       &   978.3 &   982.4 &   983.1 &   985.5 &   983.4 &   991.4 &  987.1 \\
$\omega_3$       &  2253.1 &  2259.6 &  2227.9 &  2254.7 &  2266.5 &  2275.4 & 2268.3 \\
$\omega_4$       &   925.8 &   933.6 &   932.5 &   933.8 &   930.8 &   937.2 &  935.3 \\
$\nu_1$          &  2167.0 &  2175.6 &  2154.8 &  2174.3 &  2185.0 &  & \\
$\nu_2$          &   965.6 &   970.0 &   964.9 &   969.0 &   968.3 &  & \\
$\nu_3$          &  2173.1 &  2181.0 &  2155.3 &  2175.1 &  2185.2 &  & \\ 
$\nu_4$          &   912.2 &   920.3 &   913.6 &   917.9 &   915.1 &  & \\
$\omega_1-\nu_1$ &   74.96 &   73.65 &   70.90 &   75.96 &   77.73 &    &   \\
$\omega_2-\nu_2$ &   12.77 &   12.43 &   18.14 &   16.51 &   15.10 &    &   \\
$\omega_3-\nu_3$ &   80.04 &   78.54 &   72.58 &   79.60 &   81.25 &    &   \\
$\omega_4-\nu_4$ &   13.65 &   13.34 &   18.96 &   15.89 &   15.70 &    &   \\
\end{tabular}

The CCSD(T) electron correlation method has been used throughout.

\end{table}

\begin{table}
\caption{Comparison of computed and observed fundamentals (cm$^{-1}$) 
for isotopomers of silane.} 
\squeezetable
\begin{tabular}{lccccccc}
  $i$  & $\nu_i$ & $\omega_i$ & $\nu_i$ & $\omega_i-\nu_i$ & $\nu_i$\\
       & CCSD(T)/ & best      &   best  &  best            & Expt.\\
	   & cc-pVQZ+1 & adjusted & adjusted & adjusted &\\
\hline
\multicolumn{6}{c}{$^{28}$SiH$_4$}\\
  1  &  2185.0  &  2264.2& 2186.9  &  77.34  &   2186.873254(80) \cite{a}\\
  2  &   968.3  &   986.0&  970.9  &  15.10  &    970.93451(6) \cite{b}\\
  3  &  2185.2  &  2270.1& 2189.2  &  80.84  &   2189.189680(66) \cite{a}\\
  4  &   915.1  &   929.1&  913.5  &  15.68  &    913.46871(4) \cite{b}\\
\multicolumn{6}{c}{$^{29}$SiH$_4$}\\
  1  &  2184.9  &  2264.2& 2186.8  &  77.39  & 2186.8281(5)    \cite{a} \\
  2  &   968.3  &   986.0&  970.9  &  15.09  &  970.94856(22)  \cite{b} \\
  3  &  2183.7  &  2268.4& 2187.7  &  80.70  & 2187.6494(1)    \cite{a} \\
  4  &   913.8  &   927.8&  912.2  &  15.63  &  912.18278(8)   \cite{b}\\
\multicolumn{6}{c}{$^{30}$SiH$_4$}\\
  1  &  2184.9  &  2264.2& 2186.8  &  77.43  & 2186.7855(6)    \cite{a} \\
  2  &   968.3  &   986.0&  971.0  &  15.08  &  970.95790(110) \cite{b}\\
  3  &  2182.2  &  2266.8& 2186.2  &  80.57  & 2186.1963(1)    \cite{a}\\
  4  &   912.6  &   926.6&  911.0  &  15.59  &  910.97921(12)  \cite{b}\\
\multicolumn{6}{c}{$^{28}$SiH$_3$D}\\
  1  &  2184.9  &  2265.7& 2187.4  &  78.38  &2187.40066(5) \cite{c}\\
  2  &  1590.7  &  1630.9& 1592.8(a) & 38.08(a) &1593.9595(10) \cite{d}\\
  3  &   914.5  &   928.5&  912.9  &  15.60  &912.991(1) \cite{e}\\
  4  &  2184.4  &  2270.0& 2188.4  &  81.51  &2188.50418(4) \cite{c}\\
  5  &   949.2  &   966.1&  950.6  &  15.48  &950.576(1) \cite{e}\\
  6  &   784.6  &   795.4&  784.2  &  11.29  &784.324(1) \cite{e}\\
\multicolumn{6}{c}{$^{28}$SiH$_2$D$_2$}\\
  1  & 2184.3   & 2269.9 &2187.7   & 82.19   &   2189 \cite{f}\\
  2  & 1579.4   & 1621.1 &1581.3   & 39.83   &   1587 \cite{f} (b) \\
  3  &  942.0   &  958.4 & 942.7   & 15.69   &    942.74106(4) \cite{thiel}\\
  4  &  681.3   &  689.6 & 681.3   &  8.29   &    681.62394(3) \cite{thiel}\\
  5  &  840.1   &  854.0 & 842.3   & 11.71   &    842.38121(9) \cite{thiel}\\
  6  & 1597.2   &  1640.6& 1599.7  &  40.87  &    1601 \cite{f} \\
  7  &  861.1   &  873.4 & 859.5   & 13.84   &    859.750104(4) \cite{thiel}\\
  8  & 2183.6   & 2267.2 &2187.3   & 79.81   &   2183 \cite{f} ??\\
  9  &  743.7   &  753.0 & 742.4   & 10.65   &    742.64029(3) \cite{thiel}\\
\multicolumn{6}{c}{$^{28}$SiHD$_3$}\\
  1  & 2183.1   & 2268.5 &2186.6   & 81.90   & 2187.2070(10) \cite{g}\\
  2  & 1570.8   & 1611.4 &1572.4   & 39.01   & 1573 \cite{f}\\
  3  &  676.4   &  683.9 & 675.2   &  8.72   & [682] \cite{f} (c) \\
  4  & 1596.0   & 1640.5 &1598.7   & 41.77   & 1598 \cite{f}\\
  5  &  850.2   &  863.4 & 850.6   & 12.84   &  850.680823(10) \cite{h}\\
  6  &  682.1   &  690.6 & 682.4   &  8.14   &  683 \cite{f}\\
\multicolumn{6}{c}{$^{28}$SiD$_4$}\\
  1  & 1562.6   & 1601.7 &1563.8   & 37.84   &  1563.2(10) \cite{i}\\
  2  &  687.9   &  697.5 & 689.8   &  7.75   &   685.2(2) \cite{i}\\
  3  & 1595.2   & 1640.5 &1598.0   & 42.48   &  1598.44919(43) \cite{j},1598.45(5) \cite{i}\\
  4  &  675.3   &  682.6 & 674.1   &  8.48   &   674.2(15) \cite{i}\\
\end{tabular}

(a) If Fermi resonance $2\nu_8\approx\nu_2$ is accounted for
($2\nu_8^*$=1563.9 cm$^{-1}$, $k_{288}$=21.393 cm$^{-1}$,
$\nu_2^*$=1587.1 cm$^{-1}$) we obtain $\nu_2$=1594.6 cm$^{-1}$,
and $2\nu_8$=1556.5 cm$^{-1}$.

(b)  in fact doublet at 1584 and 1591 cm$^{-1}$; we suggest 
assignment of 1584 cm$^{-1}$ to $\nu_{2}$ and of 1591 cm$^{-1}$ to 
possibly $\nu_5+\nu_9$

(c) not observed; valence force field estimate. Authors of Ref.\cite{f} speculate
that it coincides with the 683 cm$^{-1}$ band.

\end{table}

\begin{table}
\caption{Quadratic, cubic and quartic force 
constants (aJ/\AA$^m$radian$^n$) for SiH$_4$}
\begin{tabular}{lrlrlr}
$F_{11}$ & 3.04428 & $F_{22}$ & 0.41777 & $F_{44}$ & 2.92753 \\
$F_{74}$ & -0.08914 & $F_{77}$ & 0.51105 & $F_{111}$ & -6.72559 \\
$F_{221}$ & -0.16483 & $F_{441}$ & -6.55397 & $F_{741}$ & 0.05353 \\
$F_{771}$ & -0.14909 & $F_{222}$ & -0.02902 & $F_{662}$ & -0.11954 \\
$F_{962}$ & 0.13392 & $F_{992}$ & -0.29002 & $F_{654}$ & -6.51514 \\
$F_{954}$ & -0.07640 & $F_{984}$ & -0.03587 & $F_{987}$ & 0.41927 \\
$F_{1111}$ & 12.58584 & $F_{2211}$ & -0.00037 & $F_{4411}$ & 12.80551 \\
$F_{7411}$ & 0.05928 & $F_{7711}$ & -0.06795 & $F_{2221}$ & -0.01703 \\
$F_{6621}$ & 0.04882 & $F_{9621}$ & -0.00005 & $F_{9921}$ & 0.09065 \\
$F_{6541}$ & 12.78806 & $F_{9541}$ & 0.02617 & $F_{9841}$ & 0.05749 \\
$F_{9871}$ & -0.11162 & $F_{2222}$ & 0.16525 & $F_{6622}$ & -0.10547 \\
$F_{6633}$ & -0.22839 & $F_{9622}$ & -0.08371 & $F_{9633}$ & -0.02617 \\
$F_{9922}$ & 0.17090 & $F_{9933}$ & 0.30175 & $F_{9542}$ & 0.18309 \\
$F_{8762}$ & -0.04530 & $F_{4444}$ & 13.10285 & $F_{5544}$ & 12.93973 \\
$F_{7444}$ & 0.03663 & $F_{8544}$ & 0.05928 & $F_{7744}$ & 0.00877 \\
$F_{8754}$ & -0.00357 & $F_{8844}$ & -0.21095 & $F_{7774}$ & -0.24594 \\
$F_{8874}$ & -0.08029 & $F_{7777}$ & 0.26859 & $F_{8877}$ & 0.75864 \\
\end{tabular}
\end{table}

\end{document}